%% file: article.tex
\documentclass[a4paper]{article}
\usepackage[T1]{fontenc}
\usepackage[utf8]{inputenc}
\usepackage{amsmath,amssymb}
\usepackage{tikz}
\usepackage{listings}
\usepackage{graphicx}
\usepackage{amsthm}
\usepackage{multirow}
\usepackage{booktabs}
\RequirePackage[noline,noend]{algorithm2e}
\usepackage{authblk}

\theoremstyle{definition}
\newtheorem{defn}{Definition}

\usepackage{mathtools}
\DeclarePairedDelimiter{\ceil}{\lceil}{\rceil}
\DeclarePairedDelimiter\floor{\lfloor}{\rfloor}

\newcommand\auxfun[1]{\expandafter\newcommand\csname #1\endcsname{%
 \mathop{\hbox{$\mathsf{#1}$}}\nolimits}}
 \newcommand\af[1]{\mathop{\hbox{$\mathsf{#1}$}}\nolimits}

\let\mod=\relax
\auxfun{mod}

\title{Age-Partitioned Bloom Filters}
\author[1]{Ariel Shtul}
\author[2]{Carlos Baquero}
\author[3]{Paulo Sérgio Almeida}
\affil[1]{Redis Labs}
\affil[2,3]{INESC TEC and University of Minho}
\date{}
\begin{document}
\maketitle

\begin{abstract}
  Bloom filters (BF) are widely used for approximate membership queries over a
  set of elements. BF variants allow removals, sets of
  unbounded size or querying a sliding window over an unbounded stream.
  However, for this last case the best current approaches are dictionary based
  (e.g., based on Cuckoo Filters or TinyTable), and it may seem that BF-based approaches will
  never be competitive to dictionary-based ones.
  In this paper we present \emph{Age-Partitioned Bloom Filters}, a BF-based
  approach for duplicate detection in sliding windows that not only
  is competitive in time-complexity, but has better space usage than current
  dictionary-based approaches (e.g., SWAMP), at the cost of some
  moderate slack. 
  APBFs retain the BF simplicity, unlike dictionary-based approaches,
  important for hardware-based implementations, and can integrate known
  improvements such as \emph{double hashing} or \emph{blocking}.
  We present an \emph{Age-Partitioned Blocked Bloom Filter} variant which can
  operate with 2–3 cache-line accesses per insertion and around 2–4 per query, even
  for high accuracy filters.
\end{abstract}


\section{Introduction}


Bloom filters \cite{DBLP:journals/cacm/Bloom70} were developed in the 1970s to provide space and time efficient representations of sets. At the cost of a tunable false positive  probability on membership queries, bloom filters only require a small number of bits per element and are free from false negatives in queries. 
Typically a filter needs to be dimensioned in advance by deciding on a maximum allowed false positive rate and the number of elements to store. 

Standard bloom filters are not adequate for unbounded streams of elements or when recently inserted elements are more significant that older ones.
In this paper we focus on the problem of storing an approximate representation of the last $w$ elements that were inserted, while efficiently forgetting older elements. 
Solutions to this problem allow the insertion of an unlimited number of elements and the filter dimension is tied to the chosen precision and window size $w$. 

Different solutions, covered in the related work, provide different properties in terms of space use efficiency, query time complexity, programming complexity, number of extra elements (slack), and behaviour under element re-insertion. This class of bloom filters is adequate for duplicate detection in streams \cite{DengR06,ShenZ08,WangS10,WeiJZFW11,LeeC11}, 
representation of approximate caches \cite{ChangLF04},
click fraud detection \cite{ZhangG08,WalgampayaKW12}, and other uses cases \cite{LiuCG13}.

Existing solutions either strive to minimize space complexity at the cost of
relying on more complex algorithms (e.g., dictionary-based techniques
\cite{NaorY15,LiuCG13,AssafBEF18}), or are simple but inefficient in space
utilization (e.g., double-buffering with active and warm-up filters
\cite{ChangLF04,Yoon10,SubramanyamGLW16}). 

In this paper we present \emph{Age-Partitioned Bloom Filters} (APBF) and one
variant \emph{Age-Partitioned Blocked Bloom Filters} (APBBF), as new solutions
to the problem. They aim to strike a balance among implementation complexity,
space efficiency properties, and time complexity of operations. Our solutions
provide the following properties: 

\begin{itemize}
  \item Simple and efficient algorithms. A
    static bloom filter relies on $k_b$ distinct hash functions to lookup bit
    matches in $k_b$ different memory slices (array of bits). An APBF simply uses $k_a \geq k_b$ hash functions to
    look for $k_a$ \emph{consecutive matches} across a wider sequence of $k_a+l$ distinct bit memory
    slices. This is tunable by a parameter $l$ that is chosen with $k_a$ to
    preserve the same false positive rate of a static filter with $k_b$ hashes.  
  \item Tunable spatial organization, that allows APBF trading off space efficiency and slack for time efficiency, while preserving a target false positive rate. Some combinations allow a number of distinct memory accesses (the main indicator of time complexity) nearly identical to classic static bloom filters, and with modest increases in accesses (within the same order of magnitude) for combinations providing lower slack. 
  \item Compatible with blocked solutions. For settings where speed is
    critical we show that APBBF, a blocked variant of APBF, has around 2--4
    distinct cache-line accesses, even for low false positive rates. This
    allows a simpler BF-based solution to compete with dictionary-based
    techniques and provide better space efficiency in settings that tolerate
    some level of slack in expiring elements. 
\end{itemize}


The paper is organized as follows: The next section introduces a refined model of window based solutions. The following, Section \ref{sec:related}, presents an extensive review and classification of related work.  
Section \ref{sec:bloom} quickly introduces bloom filters and reviews the main mathematical properties, to be followed by Section \ref{sec:apbf} that presents and evaluates APBF. Section \ref{sec:apbbf} adds blocking and introduces APBBF. Conclusion are presented in Section \ref{sec:conc}.

\input{window-problem}

\input{related-work}

\section{Background on Bloom filters}
\label{sec:bloom}

We now briefly review Bloom filters, namely the \emph{partitioned Bloom
filter} variant which we adopt here, proposed by Mullin~\cite{Mullin83}, which
partitions the bit vector into $k$ sub-vectors of same size, which we denote
by \emph{slices}, one per hash function. The filter state is thus composed of
slices $s_1,\ldots,s_k$, each being an array of $m$ bits, initially all at
$0$.  Slices are accessed by independent hash functions $h_1,\ldots,h_k$, each
uniformly mapping arbitrary elements into $m$ positions.  In practice, double
hashing can be used to achieve similar results, without the need for $k$ hash
functions, by combining just two hash
functions~\cite{DBLP:journals/rsa/KirschM08}. This scheme can also be applied to
our proposal.
Bloom filters support two basic operations:

\begin{itemize}
  \item $\mathsf{add}(e)$: set bit $s_i[h_i(e)]$ to 1, for each $i$ in $\{1, \ldots, k\}$,
  \item $\mathsf{query}(e)$: returns true iff $s_i[h_i(e)]=1$, for
    all $i$ in $\{1, \ldots, k\}$.

\end{itemize}

From this definition there are no false negatives, i.e., once an element is
added the filter always returns true when it is queried in the future.

\subsection{Fill ratio, size and false positives}
\label{sec:fill}

Following the standard practice, we define fill ratio $r$ as the ratio of set
bits to the slice size $m$. After $n$ insertions it is given by:

\[
  r = 1 - \left( 1 - \frac{1}{m} \right)^n \approx 1 - e^{-\frac{n}{m}}.
\]

The false positive rate $p$ of a filter can be calculated in terms of the fill
rate and the number of slices as:
$$p=r^k.$$

It is well known~\cite{DBLP:journals/im/BroderM03} that, for any target false
positive rate $p$, the filter usage is optimized when $r=1/2$ (for standard
Bloom filters, and also asymptotically for partitioned Bloom filters). For this ratio,
the false positive rate $p$ is determined by the number $k$ of slices,
or equivalently, hash functions. For a target false positive rate $p$ this
number can be obtained by rounding up to the nearest integer:
$$k=\ceil{\log_2(1/p)},$$
to ensure a false positive rate less or equal to $p$. Each additional slice
leads to a linear increase in filter size and an exponential increase of
overall precision.
For a desired filter capacity of $n$ elements, we can dimension the slice size
$m$ by
$$ m = \frac{n}{\ln{2}} \approx 1.442695 \times n, $$ 
which makes clear that the slice (and filter) size is linear with its capacity.

\subsection{Time complexity}
\label{sec:time}

The time complexity of \textsf{add} and \textsf{query} operations is dominated
by the number of different memory accesses, one for each slice. This
complexity is mostly invariant with the filter capacity. In more detail

\begin{description}
\item[\textsf{add}] requires $k$ distinct accesses;
\item[\textsf{query returning true}] requires $k$ distinct accesses; 
\item[\textsf{query returning false}]  requires from 1 to $k$ distinct
  accesses since any bit at 0 terminates the query. For a full filter
  ($r=1/2$), the average number of accesses is $\frac{2^k}{2^k-1}\sum_{i=1}^k
  \frac{i}{2^i}$, upper bounded by and approaching 2 as $k \rightarrow \infty$.
\end{description}

\begin{table}
\begin{center}
  \begin{tabular}{@{}lrrlrr@{}}
  \toprule
  &&&& \multicolumn2c{query accesses} \\ \cmidrule(l){5-6}
  aimed fp & k & bits/item & actual fp & query true & query false \\
  \midrule
  0.1 & 4 & 5.77 &  0.0625  & 4.00 & 1.73 \\
  0.01 & 7 & 10.09 & 0.0078125 & 7.00 & 1.94 \\
  0.001 & 10 & 14.42 &    0.0009765625 & 10.00 & 1.99 \\
  0.0001 & 14 & 20.19 &   0.0000610351 & 14.00 & 2.00  \\
  0.00001 & 17 & 24.52 &  0.0000076293 & 17.00 & 2.00 \\
  \bottomrule
\end{tabular}
\end{center}
  \caption{Metrics for Bloom filters for different maximum false positive rates.}
  \label{tab:bf}
\end{table}

In Table \ref{tab:bf} we present values for bloom filters calibrated with
different desired false positive rates. Due to rounding up when
choosing $k$ we also show the actual (lower) maximum false positive rate, assuming
that capacity was planned aiming to stop inserting at $r=1/2$. The table
also includes how many bits must be provisioned for element and the average
number of accesses per query, for both cases of either true or false result.

\section{Age-Partitioned Bloom filters}
\label{sec:apbf}

We now present our novel proposal, \emph{Age-Partitioned Bloom Filters}. It
provides a solution to the $(w, s, p, e)$-Sliding Filter problem; it exhibits
similar computational costs as classic bloom filters, with a similar number of
memory accesses; and it has a reasonable space efficiency, better than other Bloom
filter based approaches for sliding windows, being competitive with Dictionary
approaches and much simpler.

\subsection{Structure}

An APBF uses the segmented approach, in which disjoint segments are added and
retired over time, as elements are added and expire. In the basic version, which
we present now, a segment is a slice of a partitioned Bloom filter. In
Section~\ref{sec:apbbf} we generalize the idea and use a blocked Bloom filter per
segment.

An APBF, with parameters $k$, $l$, $m$, is structured as a sequence of $k +
l$ slices ($s_0$ to $s_{k+l-1}$) with $m$ bits each, and a counter $n$ of
insertions (the size of the stream $x_1, \ldots, x_n$ up to the current time).
Parameter $l$ is relevant for the filter capacity. The filter makes use of
$k+l$ independent hash functions $h_0, \ldots, h_{k+l-1}$.

On insertions the first slices $s_0$ to $s_{k-1}$ are updated. Each batch of $g$
(discussed below) insertions is a generation. Conceptually, after each
generation, i.e., each $g$ insertions, the slices age, shifting to subsequent
slices ($s_0$ becomes $s_1$ and so on), and slice $s_{k+l-1}$ is discarded.
In practice, a circular buffer of slices is used, keeping a base index that
maintains the position of slice $s_0$ in the buffer, starting at 0. The
logical shift is then performd by zeroing slice $s_{k+l-1}$, to be reused as
the new $s_0$, and decrementing the base index, modulo $k+l$.

\begin{figure}[t]
  \begin{center}
    \includegraphics{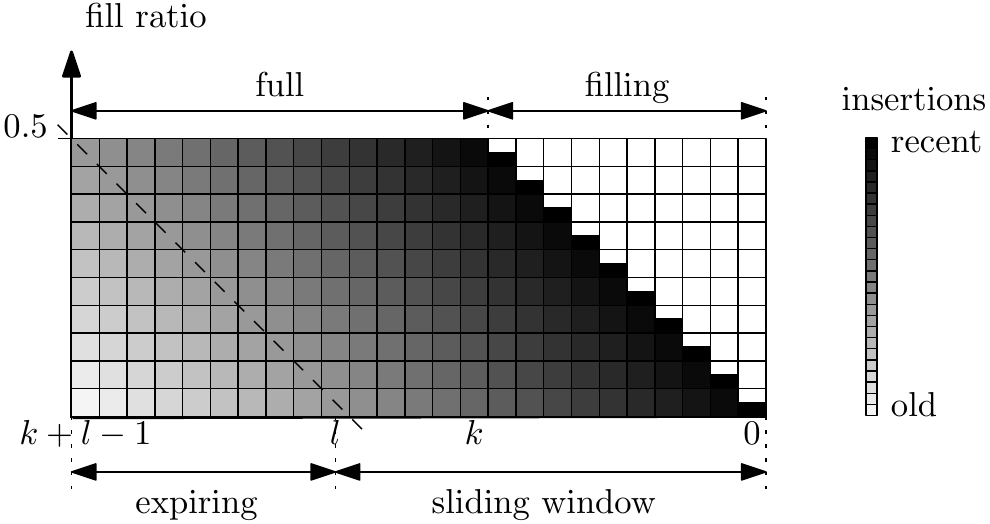}
  \end{center}
  \caption{Fill ratio of each slice from $0$ to $k+l-1$, for APBF with $k=10$
  and $l=15$. Slices from $0$ to $k-1$ are filling, while the others remain at
full capacity.}
  \label{fig:apbf-slices}
\end{figure}

Figure~\ref{fig:apbf-slices} shows how the $k+l$ slices look like in terms of
fill ratio, after some time. The first $k$ slices (with $s_0$ on the right-hand
side) are being filled with the current generation. Slices from $s_k$ onwards
stop being used for insertions, and ranges of $k$ consecutive slices, starting
from $s_0$ up to $s_l$ are checked in queries.

Even if each element is inserted by updating the first $k$ slices, i.e., the
\emph{initial} slice is always $s_0$, as generations pass, slices shift, and
the initial slice of element $x_i$ when $n$ elements have been inserted is
given by

\[ \floor{(n-1)/g} - \floor{(i-1)/g}.  \]

When querying, to check whether an element $e$ was inserted using some $k$
slices, which may have shifted meanwhile, the same $k$ hash functions used for
insertion must be used. This is achieved by having one fixed hash function per
physical bit array in the circular buffer, regardless of which logical slice is
currently mapped to it. The position in the circular buffer corresponding to
slice $s_i$ is given by

\[ \af{pos}(i) = (i - \floor{(n-1)/g}) \mod (k+l).  \]

We use $b_i$ to denote the bit array at circular buffer position $i$ and $s_i$
the logical slice mapped to $b_{\af{pos}(i)}$.
The filter uses hash function $h_i$ for bit array $b_i$ and we denote $h'_i$
for the hash function used for slice $s_i$, given by:

\[ h'_i = h_{\af{pos}(i)}. \]

In terms of the number of independent hash functions needed, it can be noticed
that two positions corresponding to slices $k$ or more indexes apart are not
used in the same insertion, and we could reuse hash functions, having only $k$
independent hash functions, but this prevented having a fixed mapping from
buffer positions to hash functions. In practice, this saving is irrelevant
if using double hashing, as the equivalent to $k+l$ hash functions can be
easily obtained.

The two filter operations, add and query, can now be defined as follows:

\begin{itemize}
  \item
$\mathsf{add}(e)$:
    if $(n \mod g) = 0$ perform a shift;
    increment $n$;
    set bit $s_i[h'_i(e)]$ to 1, for all $i$, with $0 \leq i < k$;
  \item
$\mathsf{query}(e)$: returns true iff there is a $j$, with $0 \leq j
\leq l$, such that for all $i$, with $j \leq i < j+k$, we have $s_i[h'_i(e)]=1$,
otherwise return false.
\end{itemize}

In Section~\ref{sec:query-algo} we will present an algorithm for this specification of query.

\subsection{Generations, window size and slack metrics}

Shifting one position leads to an overlap between slices used for the previous
and the current generation. The overlap only ceases when the slices are from
$k$ or more generations apart. From the point of view of an individual bit
array lifetime, it is used for insertions for $k$ generations (Figure \ref{fig:apbf-slices}), until becoming
used only for queries. This means that the generation size $g$ is the slice
capacity divided by $k$. Aiming to stop inserting when the fill ratio is
$1/2$, we have then:

\[ g = \floor{\frac{m \times \ln 2}{k}} \]


\begin{figure}
  \centering
  \begin{small}
\begin{tikzpicture}
  \matrix
  [
  ]
  {
    \node {-}; &  \node {-}; & \node {-}; & \node {-}; & \node {-}; &
    \node {\qquad\qquad}; &
    \node {-}; &  \node {-}; & \node {-}; & \node {-}; & \node {-}; &
    \node {\qquad\qquad}; &
    \node {-}; &  \node {-}; & \node[draw] {C}; & \node {-}; & \node {-}; \\[2mm]

    \node {-}; &  \node {-}; & \node {-}; & \node {-}; & \node {-}; &
    \node {\qquad\qquad}; &
    \node {-}; &  \node {-}; & \node[draw] {B}; & \node[draw] {B}; & \node {-}; &
    \node {\qquad\qquad}; &
    \node {-}; &  \node[draw] {B}; & \node[draw] {B}; & \node[draw] {C}; & \node {-}; \\[2mm]

    \node {-}; &  \node {-}; & \node[draw] {A}; & \node[draw] {A}; & \node[draw] {A}; &
    \node {\qquad\qquad}; &
    \node {-}; &  \node[draw] {A}; & \node[draw] {A}; & \node[draw] {A}; & \node[draw] {B}; & 
    \node {\qquad\qquad}; &
    \node[draw] {A}; &  \node[draw] {A}; & \node[draw] {A}; & \node[draw] {B}; & \node[draw] {C}; \\

    \node (al) {~4~~}; & \node (am) {~3~~}; & \node (an) {~2~~}; & \node{~1~~}; & \node (ar) {~0~~}; &
    \node {\qquad\qquad}; &
    \node (bl) {~4~~}; & \node (bm) {~3~~}; & \node (bn) {~2~~}; & \node{~1~~}; & \node (br) {~0~~}; &
    \node {\qquad\qquad}; &
    \node (cl) {~4~~}; & \node (cm) {~3~~}; & \node (cn) {~2~~}; & \node{~1~~}; & \node (cr) {~0~~}; \\[4mm]

    \node {}; & \node {}; & \node {\small (a)}; & \node {}; & \node {}; &
    \node {\qquad\qquad}; &
    \node {}; & \node {}; & \node {\small (b)}; & \node {}; & \node {}; &
    \node {\qquad\qquad}; &
    \node {}; & \node {}; & \node {\small (c)}; & \node {}; & \node  {}; \\[6mm] 

    \node {-}; &  \node[draw] {C}; & \node[draw] {D}; & \node {-}; & \node {-}; &
    \node {\qquad\qquad}; &
    \node[draw] {C}; &  \node[draw] {D}; & \node[draw] {E}; & \node {-}; & \node {-}; &
    \node {\qquad\qquad}; &
    \node[draw] {D}; &  \node[draw] {E}; & \node {-}; & \node {-}; & \node {-}; \\[2mm]

    \node[draw] {B}; &  \node[draw] {B}; & \node[draw] {C}; & \node[draw] {D}; & \node {-}; &
    \node {\qquad\qquad}; &
    \node[draw] {B}; &  \node[draw] {C}; & \node[draw] {D}; & \node[draw] {E}; & \node {-}; &
    \node {\qquad\qquad}; &
    \node[draw] {C}; &  \node[draw] {D}; & \node[draw] {E}; & \node {-}; & \node {-}; \\[2mm]

    \node[draw] {A}; &  \node[draw] {A}; & \node[draw] {B}; & \node[draw] {C}; & \node[draw] {D}; &
    \node {\qquad\qquad}; &
    \node[draw] {A}; &  \node[draw] {B}; & \node[draw] {C}; & \node[draw] {D}; & \node[draw] {E}; & 
    \node {\qquad\qquad}; &
    \node[draw] {B}; &  \node[draw] {C}; & \node[draw] {D}; & \node[draw] {E}; & \node {-}; \\

    \node (dl) {4}; & \node (dm) {3}; & \node (dn) {2}; & \node{1}; & \node (dr) {0}; &
    \node {\qquad\qquad}; &
    \node (el) {4}; & \node (em) {3}; & \node (en) {2}; & \node{1}; & \node (er) {0}; &
    \node {\qquad\qquad}; &
    \node (fl) {4}; & \node (fm) {3}; & \node (fn) {2}; & \node{1}; & \node (fr) {0}; \\[4mm] 

    \node {}; & \node {}; & \node {\small (d)}; & \node {}; & \node {}; &
    \node {\qquad\qquad}; &
    \node {}; & \node {}; & \node {\small (e)}; & \node {}; & \node {}; &
    \node {\qquad\qquad}; &
    \node {}; & \node {}; & \node {\small (f)}; & \node {}; & \node  {}; \\[2mm] 
  };
  \draw [<->,thick] (an.south) -- (ar.south)  node [below,midway] {k};
  \draw [<->,thick] (al.south) -- (am.south)  node [below,midway] {l};
  \draw [<->,thick] (bn.south) -- (br.south)  node [below,midway] {k};
  \draw [<->,thick] (bl.south) -- (bm.south)  node [below,midway] {l};
  \draw [<->,thick] (cn.south) -- (cr.south)  node [below,midway] {k};
  \draw [<->,thick] (cl.south) -- (cm.south)  node [below,midway] {l};
  \draw [<->,thick] (dn.south) -- (dr.south)  node [below,midway] {k};
  \draw [<->,thick] (dl.south) -- (dm.south)  node [below,midway] {l};
  \draw [<->,thick] (en.south) -- (er.south)  node [below,midway] {k};
  \draw [<->,thick] (el.south) -- (em.south)  node [below,midway] {l};
  \draw [<->,thick] (fn.south) -- (fr.south)  node [below,midway] {k};
  \draw [<->,thick] (fl.south) -- (fm.south)  node [below,midway] {l};
\end{tikzpicture}
  \end{small}
\caption{Filter evolution when adding elements and shifting. Each letter (A,B,\ldots) represents a generation of $g$ elements. With $k=3$ and $l=2$}
\label{fig:shift}
\end{figure}

In Figure \ref{fig:shift} we illustrate this process, for a filter with
$k=3$ and $l=2$. With $k=3$, each generation takes $1/3$ of the slice
capacity. The first generation $A$ is added to slices $0$ to $2$. Then we keep
shifting and adding generations $B$ and $C$. In phase (d) we see that generation
$D$ no longer overlaps with the slices used for the first generation $A$, since
we already shifted 3 times, and that one slice from generation $A$ has already
been discarded, possibly leading to elements from $A$ being reported as
missing. This because generation $A$ is no longer part of the sliding
window, but is expiring, in the transition zone, with elements having a
considerable probability of being reported as present.

To obtain the sliding window size $w$, consider the situation immediately
after $l$ shifts of the $g$ elements of generation $A$ initially inserted:
they are now in slices $l$ to $l+k-1$, we have two generations stored ($A$,
and $B$) and we can still add generation $C$ before shifting.  If we insert
generation $C$, before the next shift we have $3g$ elements stored, but
just after the shift we go back two $2g$ elements, as it will discard the last
slice and invalidate generation $A$. This means that the sliding window size
(the number of elements guaranteed to be reported with no false negative) is
two generations ($2g$). This is also seen in phase (f), after shifting
generation $E$. In general:
\[ w = l \times g, \]
and the number of elements reported with no false negatives oscilates between
$l \times g$ and $(l+1) \times g$.
Considering the filter just before a shift, we can see that the slack is $k$
generations:
\[ s = k \times g. \]
Which means that the normalized slack, relative to window size $w$ is given by
the ratio between $k$ and $l$.

\begin{figure}[t]
  \begin{center}
    \includegraphics{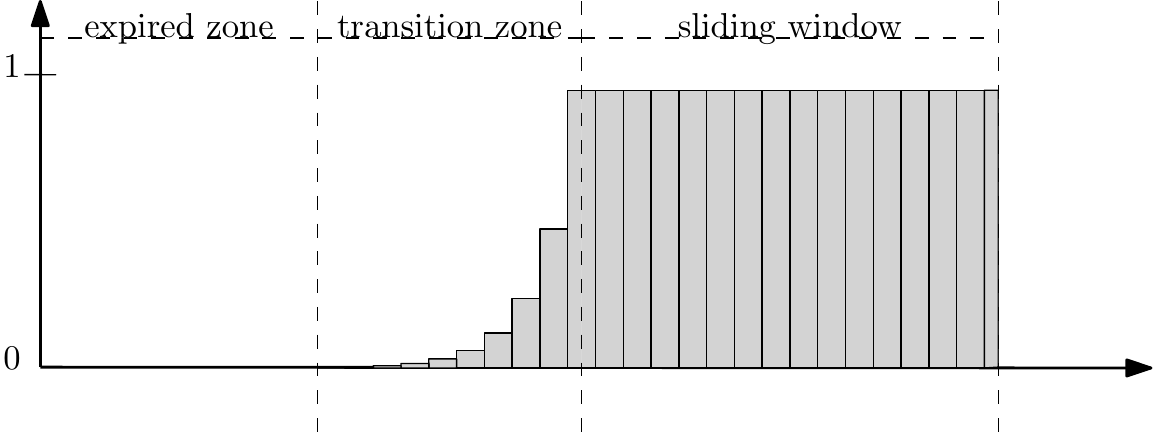}
  \end{center}
  \caption{Exponential decay in the probability of reporting in the transition zone, for APBF with $k=10$
  and $l=15$.}
  \label{fig:apbf-window}
\end{figure}

As we discussed, a more interesting metric is
the normalized probability-weighted slack (NPWS). To calculate it, or more
concretely, the worst case peak NPWS, consider again the situation just
before a shift. For the $k$ generations in the slack: there is one generation
that will be always reported, another is missing one slice and will be
reported with $1/2$ probability, the next with $1/4$ probability and so on.
Figure \ref{fig:apbf-window} illustrates the exponential decay of reported
elements from expiring generations. The peak NPWS is therefore:
\[
  \textrm{NPWS} = \frac{\sum_{i=0}^{k-1} \frac{1}{2^i}}{l} \lesssim
  \frac{2}{l},
\]
i.e., almost $2/l$ for typical $k$.

%

\subsection{False positive rate}

Calculating the false positive rate of an APBF is considerably more complex
than for a normal Bloom filter. Two factors are at play: (1) any consecutive
$k$ positive matches among the $k+l$ slices allow returning true; (2) each
match depends on the slice fill ratio which, due to shifting, are not all identical. 
After $k+l$ shifts the filter is in a steady state, storing between $l$ and
$l+1$ generations. In the worst case, just before a shift occurs, the expected
fill ratios are:

$$
[r_0, r_1, \ldots, r_{k-1}, r_{k}, \ldots, r_{k+l-1}] =
[\frac{1}{2k}, \frac{2}{2k}\ldots,\frac{k}{2k},\frac{1}{2},\ldots,\frac{1}{2}]
$$

The first $k$ slides show a linear growing gradient up to $1/2$ and the
remaining ones are all filled with that maximum fill rate. 
Given this, we will not attempt to find a closed formula for the
false-positive rate, but will derive a recursive formula to allow building a
table of useful combinations of $k$ and $l$.

In a sequence of $k+l$ slices with fill ratios $r_0, r_1, \ldots, r_{k+l-1}$,
the probability of completing a sequence of at least $k$ consecutive matches,
when starting from slice $i$, after having already $a$ consecutive matches
just before slice $i$,
considering that at slice $i$ there is $r_i$ chance of matching, is given by:

\[
  F(k, l, a, i) = 
  \begin{cases}
    1 & \textrm{if } a = k. \\
    0 & \textrm{if } i > l + a. \\
    r_i F(k, l, a+1, i+1) + (1 - r_i) F(k, l, 0, i+1) & \textrm{otherwise.}
  \end{cases}
\]


The false positive rate of an APBF is then given by $F(k,l,0, 0)$.

\subsection{Query algorithm}
\label{sec:query-algo}

Queries need to find $k$ consecutive slice matches to return true. A naïf
implementation would scan the slices linearly in search of a match sequence.
This process can be improved by a careful choice of starting position,
accumulation of matching sub-sequences and jumping when failing a match. 

\begin{figure}[t]
  \begin{center}
    \includegraphics{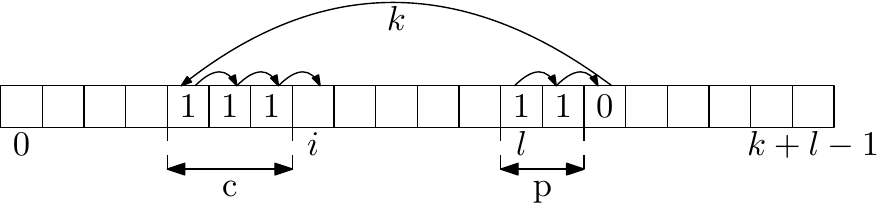}
  \end{center}
  \caption{Progress of query algorithm to find $k$ consecutive matching slices, in filter
    with $k + l$ slices, starting from slice $l$, now at slice $i$ after
  counting $c$ matches after the last $k$ slices jump, and $p$ previous
matches before the jump.}
  \label{fig:query}
\end{figure}

Figure~\ref{fig:query} illustrates how the algorithm evolves. Starting from
slice $l$, it keeps a count $c$ of consecutive matches. When a match fails it
moves to the left $k$ slices, moves counter $c$ to a previous counter $p$,
resets the counter $c$, and starts again. It stops either when the sum $p+c$
reaches $k$, or the current slice index $i$ becomes negative.
It is shown as Algorithm~\ref{algo:query}.

The algorithm starts $k$ slices from the end and jumps backwards, as it gives
better average number of memory accesses than starting from the side of the
initial slices. Similar to the false positive rate, a recursive formula for
the expected number of accesses when the query returns false is easily
obtained as $A(k, l, 0, 0, 0, l) / (1 - F(k,l,0, 0))$, with $A$ recursively
defined as:
\[
  A(k, l, p, c, a, i) = 
  \begin{cases}
    a & \textrm{if } i < 0. \\
    0 & \textrm{if } p + c = k. \\
    r_i A(k, l, p, c+1, a+1, i+1) \\
    {} + (1 - r_i) A(k, l, c, 0, a+1, i-k) & \textrm{otherwise.}
  \end{cases}
\]
where $p$ and $c$ have the same role as in the algorithm and $a$ accumulates
the number of accesses so far. Similar formulas can be obtained for the cases
of querying elements in the window or for false positives.

\begin{algorithm}[t]
    \DontPrintSemicolon
    \SetKwBlock{function}{function}{}
    \SetKw{return}{return}
    \function({$\af{query}(e)$}) {
      $i := l, \; p := 0, \; c := 0$ \;
      \While {$i \geq 0$} {
        \If {$s_i[h'_i(e)] = 1$} {
          $c := c + 1, \; i := i + 1$ \;
          \If {$p + c = k$} {
            \return $\af{true}$ \;
          }
        } \Else {
          $i := i - k, \; p := c, \; c := 0$ \;
        }
      }
      \return $\af{false}$ \;
    }
  \caption{Query algorithm.}
  \label{algo:query}
\end{algorithm}

%

\subsection{Comparison with a Bloom filter}

Having now an efficient query strategy and knowing the false positive rate of
an APBF, we are able to evaluate and compare an APBF with a Bloom filter.

First we determine the space efficiency of an APBF with respect to a Bloom
filter. Consider a Bloom filter with $k_b$ slices and an APBF with $k_a+l$
slices such that the false positive rates are identical, using the formulas
presented above, with the APBF sliding window size $w$ the same as the Bloom
filter capacity.
The relative efficiency of the APBF, i.e., the ratio between the Bloom filter
memory usage and the APBF memory usage is:

$$\mathrm{eff}=\frac{k_b}{k_a}*\frac{l}{k_a+l}$$

In Table \ref{tab:fbf} we present several metrics for APBF. For these
configurations, the space efficiency rate varies from $36\%$ to $68\%$ of the efficiency of a
Bloom filter. The worst efficiency occurs in the fast query
configurations and high overall false positive rate, and space efficiency tends
to improve for higher accuracies and for higher numbers of slices $k_a+l$. In
practice, the false positive rate is defined by the target application
requirement. If space is not critical then the choice of
$k_a+l$ should fall at or near the combination with the lower number of
slices, when $k_b=k_a$. This is specially the case for high accuracy filters,
or when NPWS is not an issue. Even then, there is no reason for $l$ being
much larger than $k$, when aiming for modest NPWS, as seen by combinations
$(k=6,\; l=14)$, or $(k=9,\; l=14)$, or $(k=12,\; l=14$).

\begin{table}
  \begin{center}
  \begin{tabular}{@{}lrrrrrrrr@{}}
  \toprule
    &&&&& \multicolumn3c{query accesses} \\ \cmidrule(lr){6-8}
    aimed fp & $k$ & $l$ & actual fp & eff & window & fp & false & \textsc{npws} \\
  \midrule
  \multirow5*{0.1}
   & 4 &  3 & 0.100586 & 0.36 &  4.71 &  4.38 &  2.16 & 0.58 \\
   & 5 &  7 & 0.101603 & 0.38 &  6.17 &  5.76 &  3.42 & 0.28 \\
   & 6 & 14 & 0.098623 & 0.39 &  8.04 &  7.58 &  5.42 & 0.14 \\
   & 7 & 28 & 0.099033 & 0.38 & 10.73 & 10.25 &  9.10 & 0.07 \\
   & 8 & 56 & 0.100234 & 0.36 & 14.87 & 14.39 & 15.60 & 0.04 \\
   \midrule
   \multirow5*{0.01}
   &  7 &  5 & 0.011232 & 0.39 &  7.81 &  7.40 & 2.02 & 0.40 \\
   &  8 &  8 & 0.010244 & 0.41 &  8.88 &  8.62 & 3.09 & 0.25 \\
   &  9 & 14 & 0.010212 & 0.45 & 10.50 &  9.89 & 3.79 & 0.14 \\
   & 10 & 25 & 0.010076 & 0.47 & 12.46 & 11.80 & 5.85 & 0.08 \\
   & 11 & 46 & 0.009948 & 0.49 & 15.24 & 14.56 & 9.55 & 0.04\\
   \midrule
   \multirow5*{0.001}
   & 10 &  7 & 0.001211 & 0.40 & 10.86 & 10.42 & 1.85 & 0.28 \\
   & 11 &  9 & 0.000918 & 0.41 & 11.89 & 11.52 & 2.15 & 0.22 \\
   & 12 & 14 & 0.000981 & 0.45 & 13.11 & 12.75 & 3.21 & 0.14 \\
   & 13 & 23 & 0.000928 & 0.50 & 14.74 & 14.06 & 4.20 & 0.09 \\
   & 14 & 40 & 0.000988 & 0.53 & 16.85 & 16.16 & 6.75 & 0.05 \\
   \midrule
   \multirow5*{0.0001}
   & 14 & 11 & 0.000099 & 0.42 & 14.91 & 14.51 & 1.93 & 0.18 \\
   & 15 & 15 & 0.000100 & 0.44 & 15.93 & 15.69 & 3.08 & 0.13 \\
   & 16 & 22 & 0.000097 & 0.48 & 17.41 & 16.84 & 3.36 & 0.09 \\
   & 17 & 36 & 0.000099 & 0.53 & 19.09 & 18.46 & 5.19 & 0.06 \\
   & 18 & 63 & 0.000099 & 0.57 & 21.56 & 20.79 & 7.68 & 0.03 \\
   \midrule
   \multirow5*{0.00001}
   & 17 & 13 & 0.000011 & 0.42 & 17.92 & 17.51 & 1.81 & 0.15 \\
   & 18 & 16 & 0.000009 & 0.44 & 18.93 & 18.62 & 2.20 & 0.12 \\
   & 19 & 22 & 0.000010 & 0.56 & 20.15 & 19.80 & 3.16 & 0.09 \\
   & 20 & 33 & 0.000010 & 0.62 & 21.70 & 20.94 & 3.68 & 0.06 \\
   & 21 & 54 & 0.000010 & 0.68 & 23.61 & 22.83 & 5.63 & 0.04 \\
   \bottomrule
\end{tabular}
  \end{center}
  \caption{Metrics for APBF with $k + l$ slices, for different aimed false
    positive rates: actual fp rate, space efficiency relative to BF, expected
    number of accesses in queries (for window items, false positives
    and reported false) and peak NPWS.}
  \label{tab:fbf}
\end{table}

With the choice of an APBF with the same $k$ as a Bloom filter, comparing
Tables \ref{tab:bf} and \ref{tab:fbf} we can observe that the number of memory
accesses in Bloom filters and APBF is roughly equivalent. An APBF has a minimal
overhead for queries that return true (either elements in the window or false
positives) and in higher precision settings they even improve on Bloom filters
when queries return false, due to the contribution of slices with fill ratio bellow $1/2$.

%
%

\subsection{Accuracy under workloads}


Together with the algorithm we developed a C implementation of \emph{Age-Partitioned Bloom Filters}. In this section we evaluate this implementation with respect to the actual false positive rate under a range of synthetic workloads.

\subsubsection{Accuracy}

\begin{figure}
  \begin{center}
  \includegraphics[width=120mm]{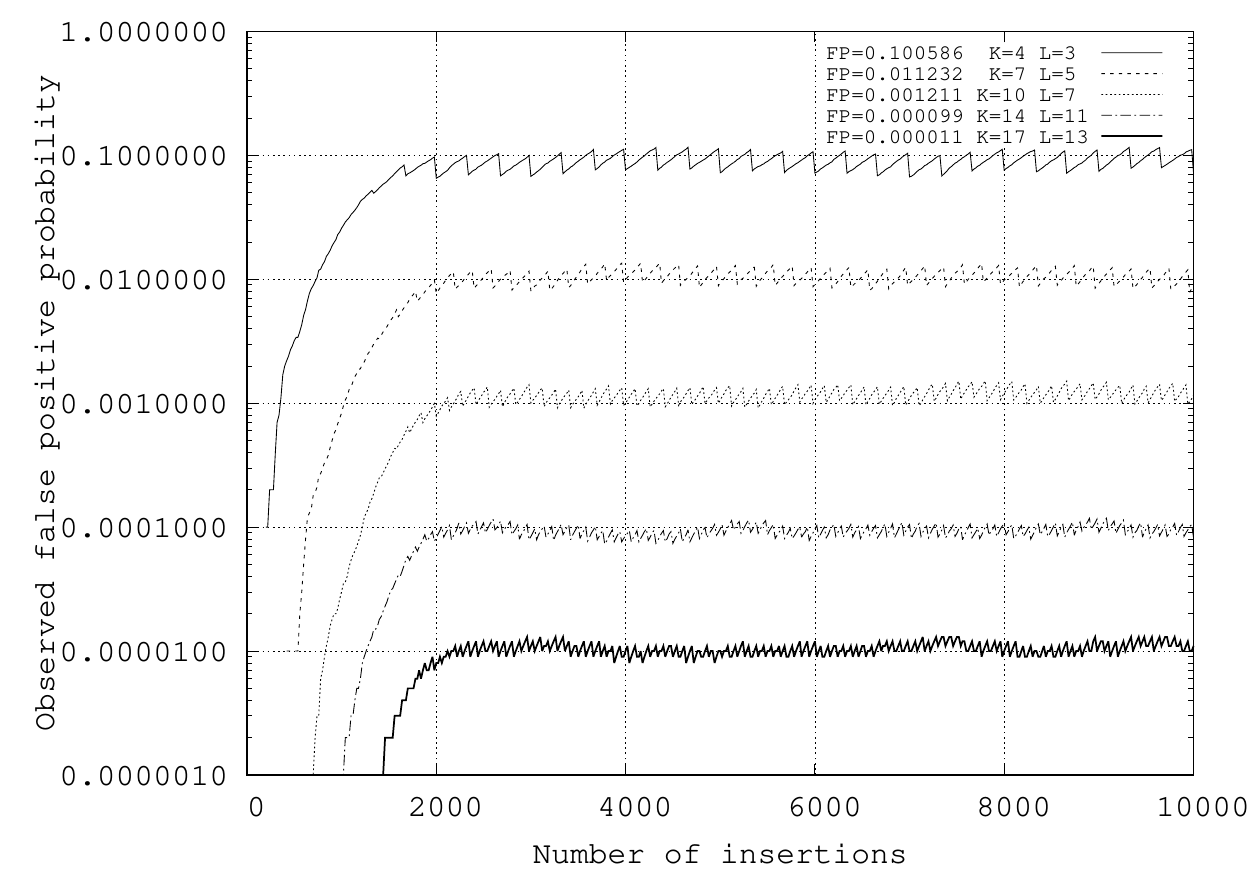}
  \end{center}
  \caption{Experimental validation of false positive rates for low $(k,l)$ age-partitioned filters. $10^4$ distinct insertions over a window $w=10^3$ elements.} 
  \label{fig:impfp}
\end{figure}

We tested for the lack of false negatives by confirming that all the last window $w$ inserted elements are reported as present. Testing for the level of false positives is more interesting and is presented on Figure \ref{fig:impfp}. Each line shows the measured false positive rates for filters with the lowest $(k,l)$ pair in Table \ref{tab:fbf}. Filters were dimensioned to hold $10^3$ elements, start empty and are subject to $10^4$ distinct insertions. 
For each filter configuration of false positive rate $1/10^i$, each sample shows the false probability outcome of probing of $10^i*10^4$ distinct elements known not to be present. This allowed increasing the precision where needed, for the more tight false positive rates. 

The figure shows that as elements are added to the filter the false positive rate stabilizes around the configured maximum rate. Particularly, in the case of the filter with $0.1$ precision we can observe a clear zig-saw effect as the filter alternates between minimum and maximum capacity. Higher $(k,l)$ pairs increasingly attenuate this effect, and the same goes for higher precisions as the figure shows. 

%
%

\subsubsection{Workloads with duplicate elements}

Our calibration of the false positive rate considered the worst case of a workload with a stream of distinct elements, since any duplicates will only decrease the fill rate and thus improve the false positive. In practical settings many workloads will exhibit duplicate elements, and this a common use case for window filters \cite{AssafBEF18,DengR06,ShenZ08,WangS10,WeiJZFW11,LeeC11}. 

\begin{figure}
  \begin{center}
  \includegraphics[width=120mm]{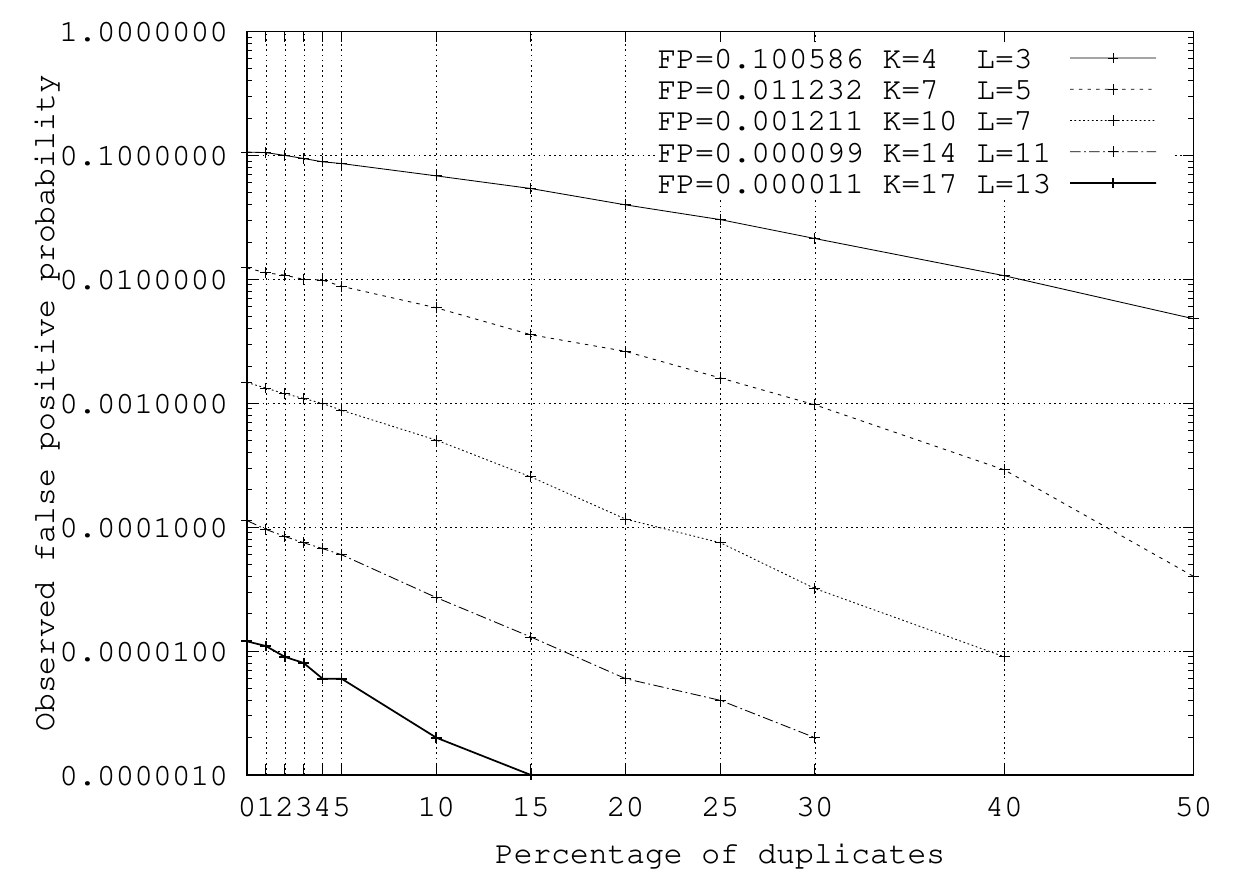}
  \end{center}
  \caption{Decay of false positives under uniform duplication. Window $w=10^3$
  elements against different duplicate rates.} 
  \label{fig:dupu}
\end{figure}

We generate synthetic workloads with different rates of duplicates, that we
define as $1-|\textrm{unique}(W)|/|W|$, where $W$ is a $w$ length sequence of
possibly non unique elements.
In Figure \ref{fig:dupu} we observe the impact of increasing the rate of
duplicates in the decrease of observed false positive rate. We see an
exponential decrease of the false positives as the duplicates increase. In
practice, if the use case ensures a known duplication rate it is possible to
adjust the calibration of the filter to a less tight false positive rate and
still have the desired false positive rate over the actual workload. 

\begin{figure}
  \begin{center}
  \includegraphics[width=120mm]{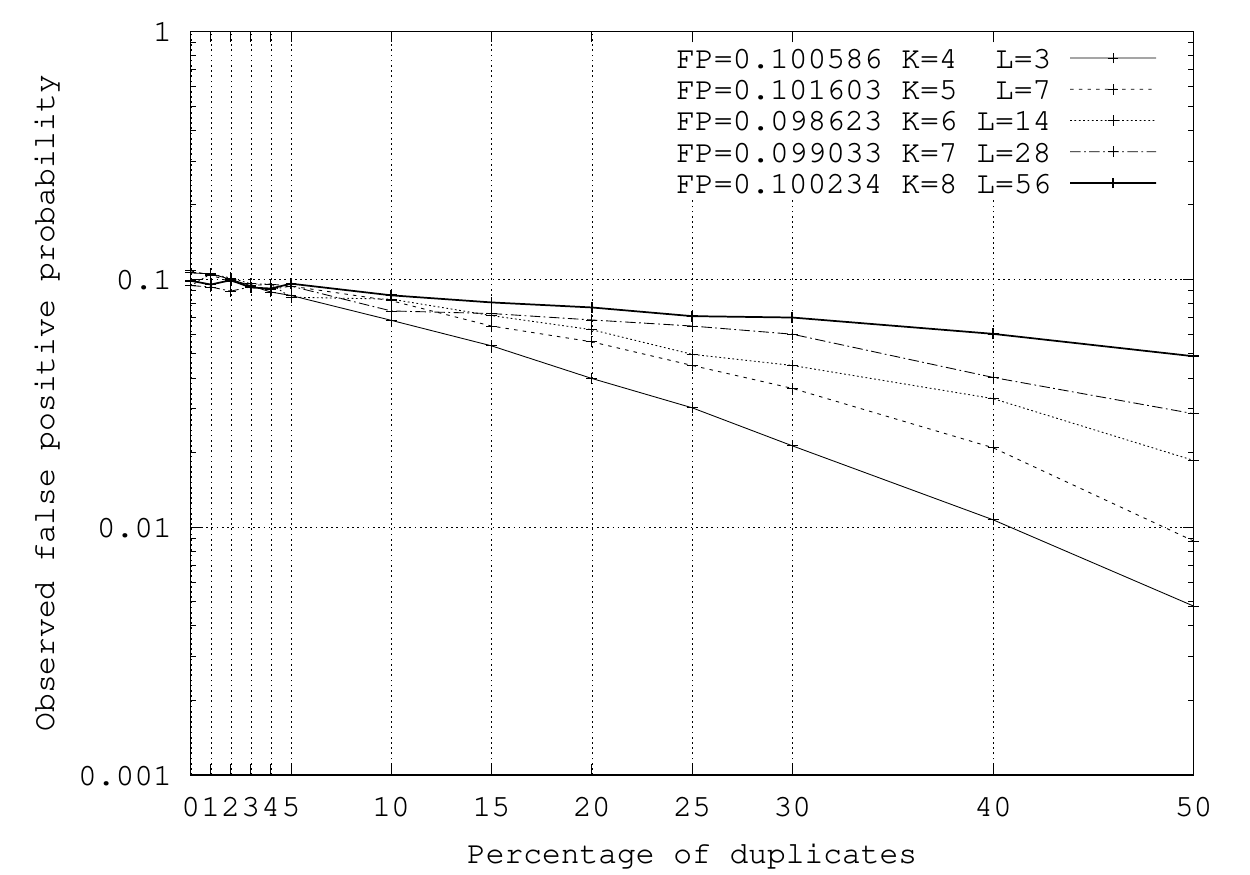}
  \end{center}
  \caption{Decay of false positives under uniform duplication. Different (k,l) configurations for 0.1 false positive rate. Window $w=10^3$ elements against different duplication rates.} 
  \label{fig:dupul}
\end{figure}

This marked decrease of false positive rate with duplicates results from the
sharing of slices when elements are inserted, since event after shifting
slices overlap for some time and are only disjoint after $k$ shifts. This
\emph{sharing effect} should be more pronounced with low $(k,l)$ pairs since
they both lead to larger slice size and to more overlap. In Figure
\ref{fig:dupul} we test different $(k,l)$ pairs having similar accuracy around
$0.1$. As expected, we observe that configurations with more slices have a
lower impact in the decrease of false positive rates for workloads with
duplicates. 

All these results considered a uniform occurrence of duplicates. If the duplicates come from a Zipfian distribution, as is common in many relevant workloads \cite{DBLP:journals/glottometrics/AdamicH02}, some elements will occur duplicated more frequently than others and the reduction of the false positive rate is expected to be even more abrupt due increased sharing.


\section{Age-Partitioned Blocked Bloom Filters}
\label{sec:apbbf}

We have shown that APBF provides a solution to the sliding filter problem while having, with low $k,l$ choices, a comparable number of distinct memory accesses when comparing with classic static bloom filters (cf. Tables \ref{tab:bf} and \ref{tab:fbf}). Blocked bloom filters~\cite{PutzeSS09} have been proposed to reduce the number of memory accesses, improving time complexity of bloom filters. The strategy is to avoid $k$ distinct memory accesses by first sharding elements into small blocks, that can fit a cache line or a processor register, and only then apply the usual $k$ hash functions to select the bits. Next we analyse how these techniques are applicable here. 

The essential idea of an APBF is to spread the hashes to segments that will be
retired at different times. The basic APBF scheme presented above uses one
slice (with one hash) of a partitioned Bloom filter per segment, i.e., as the
unit of aging/expiration, but this is naturally generalized to using any kind of Bloom
filter for each segment. All theory from the previous section still applies
(e.g., the formula for false positives, or the query algorithm) by replacing
the slice fill ratio by the subfilter false positive rate, and delegating
accesses to the subfilter.

\begin{figure}[t]
  \begin{center}
    \includegraphics{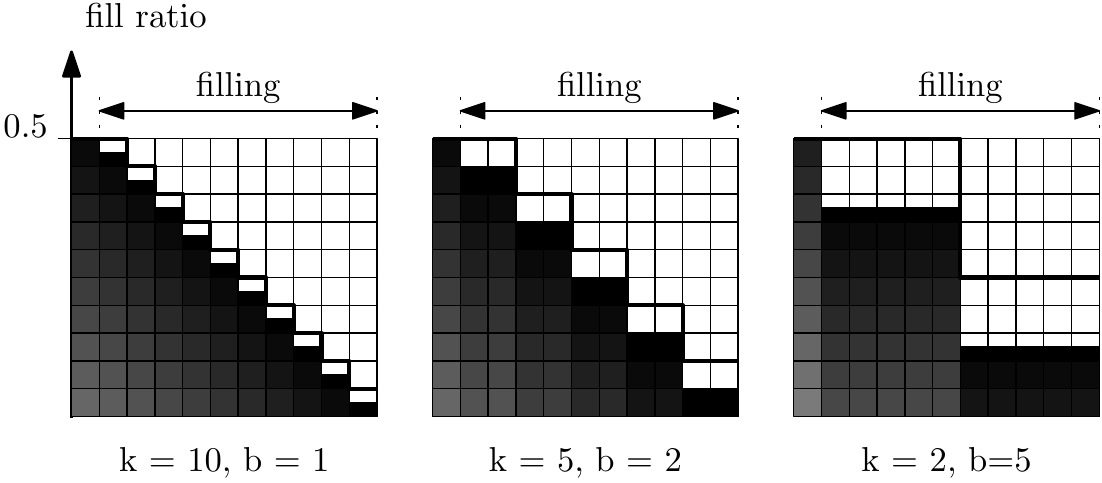}
  \end{center}
  \caption{Fill ratios for three designs: $k=10$ with one slice per segment,
    $k=5$ with two slices per segment, and $k=2$ with 5 slices per segment.}
\label{fig:bapbf-slices}
\end{figure}

One possible design is each segment being a partitioned Bloom filter with
$b$ parts, set by $b$ hash functions. In Figure~\ref{fig:bapbf-slices}, the
segments being filled of three alternative filters are shown: a basic APBF
with $k=10$ (and one slice per segment); another using $k=5$ with two slices per
segment; and one using $k=2$ and 5 slices per segment. With the same ratio
between $k$ and $l$, and using the same $m$ bits per part, the three designs
use the same total memory of $(k+l)\times b \times m$, and have the same $w$
capacity. The alternative designs will have larger generations and jumps,
being less ``smooth'': if the first design with $k=10$ has $g$ elements per
generation, the second with $k=5$ will have $2\times g$ elements per
generation, and the third will have $5\times g$.

A more interesting design, which we call \emph{Age-Partitioned Blocked Bloom
Filter} (APBBF), is using a blocked Bloom filter~\cite{PutzeSS09}
(BBF) per segment. These use a hash function to choose a $B$ sized block (e.g.,
a 512 bit cache-line or a 64 bit word) in the bit-vector, in which the $b$ bits
will be set or queried by $b$ hash functions. Even better, we can make each
block a partitioned Bloom filter, to have exactly $b$ bits set/queried. If we
make $b$ a power of two, we can simply extract $b \times \log_2(B/b)$ bits
from the result of a single hash function to choose the $b$ bits in the block
to set/query.
Because we are using $k$ filters per element insertion/query, we are
effectively using a multiblocking~\cite{PutzeSS09} scheme, which is suitable
for higher accuracy with little performance degradation over a standard Bloom
filter. An APBBF is, therefore, an age partitioned multiblocked Bloom filter
with partitioned Bloom filters for blocks. This scheme improves substantially
the weaker aspect of the basic APBF: the number of accesses, specially for
insertions or queries returning true.

\begin{table}[t]
  \begin{center}
  \begin{tabular}{@{}rrrrrrrrrrr@{}}
  \toprule
    &&&&
       \multicolumn2c{false positive rate}
    && \multicolumn3c{query accesses}
    \\ \cmidrule(lr){5-6} \cmidrule(lr){8-10}
    $k$ & $l$ & $B$ & $b$ & \textsc{apbbf} & \textsc{apbf} & cap. & win. & fp & false & \textsc{npws} \\
\midrule
2 & 3 &  64 &  4 & 0.0159865 & 0.0180539 &    0.968 &   3.30 &  2.72 &  2.25 & 0.36 \\
2 & 3 & 512 &  4 & 0.0121825 & 0.0180539 &    0.996 &   3.31 &  2.71 &  2.23 & 0.35 \\
2 & 5 &  64 &  4 & 0.0257532 & 0.0335359 &    0.968 &   3.80 &  3.28 &  3.37 & 0.21 \\
2 & 5 & 512 &  4 & 0.0197654 & 0.0335359 &    0.996 &   3.81 &  3.26 &  3.33 & 0.21 \\
\midrule
3 & 5 &  64 &  4 & 0.0016005 & 0.0017134 &    0.968 &   4.40 &  3.85 &  2.29 & 0.22 \\
3 & 5 & 512 &  4 & 0.0010688 & 0.0017134 &    0.996 &   4.40 &  3.83 &  2.26 & 0.21 \\
3 & 8 &  64 &  4 & 0.0026767 & 0.0031769 &    0.968 &   4.92 &  4.41 &  3.43 & 0.13 \\
3 & 8 & 512 &  4 & 0.0017993 & 0.0031769 &    0.996 &   4.92 &  4.39 &  3.38 & 0.13 \\
\midrule
2 & 3 &  64 &  8 & 0.0005608 & 0.0001122 &    0.936 &   3.33 &  2.70 &  2.05 & 0.34 \\
2 & 3 & 512 &  8 & 0.0000738 & 0.0001122 &    0.992 &   3.33 &  2.68 &  2.02 & 0.33 \\
2 & 5 &  64 &  8 & 0.0009231 & 0.0002342 &    0.936 &   3.80 &  3.23 &  3.08 & 0.20 \\
2 & 5 & 512 &  8 & 0.0001226 & 0.0002342 &    0.992 &   3.80 &  3.21 &  3.03 & 0.20 \\
\midrule
3 & 5 &  64 &  8 & 0.0000104 & 0.0000007 &    0.936 &   4.40 &  3.82 &  2.05 & 0.20 \\
3 & 5 & 512 &  8 & 0.0000005 & 0.0000007 &    0.992 &   4.40 &  3.78 &  2.02 & 0.20 \\
3 & 8 &  64 &  8 & 0.0000178 & 0.0000014 &    0.936 &   4.88 &  4.35 &  3.08 & 0.13 \\
3 & 8 & 512 &  8 & 0.0000009 & 0.0000014 &    0.992 &   4.88 &  4.32 &  3.03 & 0.13 \\
   \bottomrule
\end{tabular}
  \end{center}
  \caption{Metrics for APBBF with $k + l$ segments, each a BBF with $B$ block
    size and $b$ hashes per block, showing: fp rate, fp rate of an APBF with
    parameters $(k*b, l*b)$ using the same memory, relative capacity to
    such APBF, expected number of block accesses in queries (for window items,
    false positives and reported false) and peak NPWS.}
  \label{tab:apbbf}
\end{table}

In Table~\ref{tab:apbbf} we show metrics of APBBFs, for some combinations of
$k$, $l$, $B$ and $b$, from low accuracy ($k=2$ and $b=4$), moderate accuracy
($k=3$ and $b=4$ and very high accuracy ($b=8$). The table
shows how the false positive rate compares with a basic APBF using the
same memory and parameters $(k*b, l*b)$, and shows its capacity (window
size) relative to the APBF. It can be seen that while word-sized blocks are not
suitable for the $b=8$ scenarios due to the large fp degradation, 
using a word-sized block for $b=4$ is a viable choice, at some fp cost.
If slack is a not a big concern, we can make the scheme have 2--3
word/cache-line accesses per insertion and around 2--4 per query,
even for very high accuracy filters.

It should be noticed that even if slack is worse than in a basic APBF,
an APBBF with even a low $k=2$ is substantially different from a naïf
design using a sequence of BBFs: the APBBF with $(k=2,\; l=5, \; B=512,
\; b=8)$ has two block accesses for insertions and around 3--3.8 block
accesses for queries, while a naïf design with the same NPWS=0.2 would need 6
filters, and therefore at least 6 block accesses for queries, or 12 if a
multiblocked scheme with two blocks per BBF were used to avoid the fp
degradation at these high accuracies.


\paragraph{Comparison with dictionary-based techniques}

With a small number of memory accesses, APBBF overcomes the main drawback of
basic APBF and is now closer to the number of accesses that are typical in
dictionary-based techniques.
One of the best, SWAMP, is very flexible, providing other information like
number of disting elements and number of occurences, while ensuring zero
slack. Nevertheless, it uses more bits per stored element than
APBF. In Table \ref{tab:swamp} we compare the storage cost of SWAMP for a
window of $2^{16}$ elements (approximate values from Figure 3(a) in
\cite{AssafBEF18}, where SWAMP already improves over SWBF \cite{LiuCG13} and
TBF \cite{ZhangG08}) with the cost of APBBF for the same false positive rate.
We can conclude that, if we do not need all the features of SWAMP and some
slack is not a concern, our design is more compact and
considerably simpler, namely being suitable to hardware-based implementations.

\begin{table}[t]
  \begin{center}
    \begin{tabular}{@{}rrrrrrrrr@{}}
      \toprule
      &&&&& \multicolumn2c{bits per element} & \multicolumn2c{\textsc{npws}}\\ 
      \cmidrule(lr){6-7} 
      \cmidrule(lr){8-9}
      $k$ & $l$ & $B$ & $b$ & fp & \textsc{swamp} & \textsc{apbbf} & \textsc{swamp} & \textsc{apbbf}  \\
      \midrule
      2 & 5 & 512 & 4 & 0.0197654 & 32.0 & 16.2 & 0.00 & 0.21 \\
    3 & 8 & 512 & 4 & 0.0017993 & 39.8 & 23.9  & 0.00 & 0.13  \\
    2 & 5 & 512 & 8 & 0.0001226 & 48.5 & 32.6  & 0.00 & 0.20  \\
    3 & 8 & 512 & 8 &  0.0000009 & 64.6 & 48.0  & 0.00 & 0.13 \\
    \bottomrule
\end{tabular}
  \end{center}
  \caption{Comparison of APBBFs, in four previous configurations, with
  SWAMP, in terms of memory usage (bits per element) for the same false
positive rate.}
  \label{tab:swamp}
\end{table}

\section{Conclusions}
\label{sec:conc}

In this paper we have shown a new Bloom filter based approach for filters over
a sliding window. It is competitive time-wise and better space-wise
than current approaches, both Bloom filter and dictionary based.
Our design is a segmented-based approach, inspired by partitioned Bloom
filters, that spreads hashes over different segments, that will be retired
at different times. We use an efficient algorithm to look up for $k$
consecutive matches along a configurable larger number of $k+l$ segments.
This allows decaying expired entries at the granularity of segments, each with
a subset of all hashes.

Unlike current segmentation-based ones, it does not waste space resulting
from using the same hashes to set bits in different segments. It also avoids
the use of counters/timestamps that are typical of both BF-based and
some dictionary-based solutions, improving space complexity.
We derived recursive formulas for false positive rates, and number of accesses
and verified that they match the empiric results from experiments. We also
studied how the presence of duplicates in the stream benefits false positives.

Finally, we have overcome the main drawback of basic APBF -- number of memory
accesses -- by presenting a blocked variant, the Age-Partitioned Blocked Bloom
Filters. If having some slack is not an issue, APBBFs are currently, as far as
we now, the most space efficient approach, while being time-wise competive to
the best approaches. Their simplicity is also important to hardware-based
designs, unlike dictionary-based approaches.



\bibliographystyle{plain}
\bibliography{bib,streaming-bloom}

\end{document}

%% file: window-problem.tex
\section{Window models and sliding filters}

One of the first papers~\cite{MetwallyAA05} that proposes a solution based on
Bloom filters to the problem of duplicate detection in streams also
revisits window models for streams, distinguishing the \emph{Landmark Window},
\emph{Sliding Window} and \emph{Jumping Window}.

A Landmark Window considers the elements that have occurred since a specific
landmark, being mostly time-based; e.g., elements in the current day or week.
For this a simple reset is enough upon a landmark occurring, which means that
any data-structure can be used as long as it can scale to hold elements until
the next landmark.

Streaming-dedicated approaches either address the sliding window model,
focusing on the last $N$ elements, or the jumping window model, where elements
are grouped in generations, representing non-overlapping sub-windows. This is
a compromise (compared with the more precise sliding window model) allowing
summarizing each generation individually, or using course granularity
generation-based counters (as opposed to individual element-age counters).

Most jumping window based approaches are not time based, as they rely on the
capacity of components, using the element position in the stream as criteria
to group elements in generations. Some could easily be made time-based by
using scalable sub-structures.

For probabilistic approaches~\cite{NaorY15} extends the sliding window model to
allow a \emph{slack} parameter: how many inserted elements that just left the
sliding window can be reported (as present) or considered absent. For window $w$ and
slack $s$, the more recent $w$ elements must be always reported as present,
the next $s$ elements can be either reported as present or considered absent,
and the elements before the more recent $w + s$ elements should be ideally (if
there were no false-positives) considered absent.

\subsection{A novel extension to the sliding window model}

Our first small contribution is to extend the sliding window model for
probabilistic queries by introducing a variant of the slack parameter. It aims
to better describe the relationship between the overly reported elements that
just left the window and the window size.

The slack parameter from~\cite{NaorY15} refers to how many elements that just
left the window can be reported, regardless of the probability of such
reporting. As an example, this parameter does not distinguish two
filters with the same slack $s=100$, but where one always reports those 100
elements from another that will report an average of 10 of those 100 elements.
Depending on the application scenario, it could even be preferable a filter
with a slack 200 but which reports an average of 20 to a filter with a slack
of 100, reporting always those 100 elements.

We introduce what we call \emph{probability-weighted slack} (PWS): the average
number of elements that just left the window but will be reported as present.

\begin{defn}[Probability-weighted slack]
Given a filter $f$ for sliding window $w$ and slack $s$, the probability-weighted
slack (PWS) is the number of the $s$ elements before the more recent
$w$ elements that will reported in average by filter $f$.
\end{defn}

Even more interesting as a filter configuration parameter will be its variant
relative to the window size: the \emph{normalized probability-weighted slack}
(NPWS), which is the PWS divided by the window size.

\begin{defn}[Normalized probability-weighted slack]
  Given filter $f$ for sliding window $w$ with PWS $p$, the
normalized probability-weighted slack is $p / w$.
\end{defn}

\begin{figure}[t]
  \begin{center}
    \includegraphics{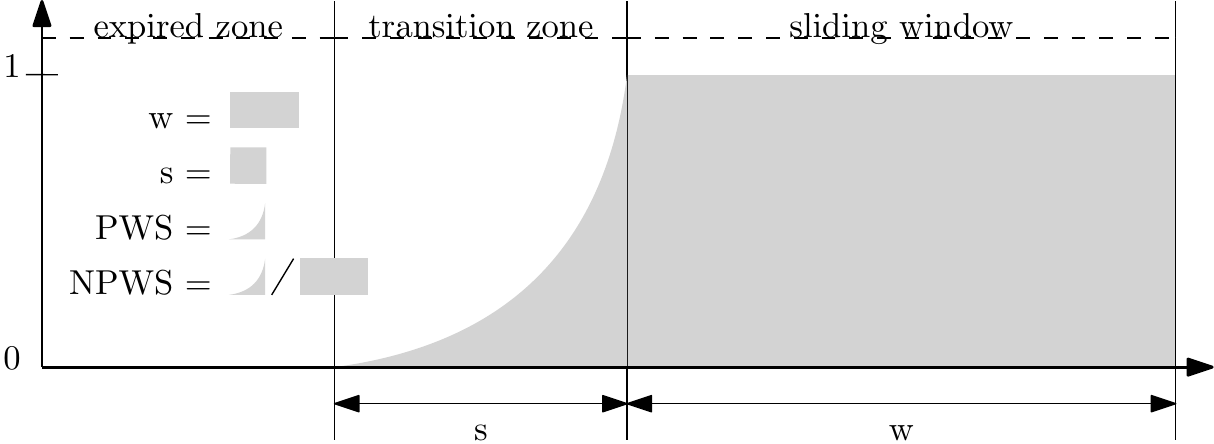}
  \end{center}
  \caption{Probability of reporting in a sliding window of size $w$, with
    slack $s$, illustrating the sliding window zone where elements are always
    reported, the transition zone where elements are sometimes reported, and
  the expired zone where elements should be considered not present.}
  \label{window-model}
\end{figure}

E.g., if we say that a filter has NPWS=10\%, for a window size $N = 1000$ it
means that we expect to have 100 element in the transition zone defined by the
slack to be reported as present (even if we do not know exactly which ones).
This PWS or NPWS parameter can be used together with the slack and window size
to better characterize the windowing behavior.
Figure~\ref{window-model} illustrates these concepts for a filter which
reports elements outside the sliding window with a probability which
approaches 0 (stopping at a very small false-positive rate) as elements age.
Our proposed solution below has such a behavior, having an exponential
decaying in the transition zone.

The PWS definition above just uses ``average'', i.e., average over the filter
lifetime. For filters which group elements and have ``jumps'', it may be more
appropriate to consider the ensemble average conditioned on the current phase
of the filter life-cycle. This defines a ``current'' PWS, which may vary along
time. Then, it is possible to talk about \emph{peak PWS}, more appropriate as
a worst-case PWS to use as criteria for such filters.

\subsection{The $(w, s, p, e)$-Sliding Filter problem definition}
\label{sec:slide}

Considering the extended sliding window model with (N)PWS above, we now define the
conditions which a filter, such as our proposal below, must respect.
Given an unlimited stream of elements $x_1, x_2, \ldots$ from some universe, and
parameters $w$ (sliding window size), $s$ (slack), $p$ (probability-weighted
slack), and $e$ (false-positive rate), we want a filter to represent
approximately the more recent elements of the stream. The elements are given
one at a time, with an arrival being signaled to the algorithm through the
$\mathsf{add}$ operation. Queries about presence can be performed through the
$\mathsf{query}$ function, which returns true of false. Unlike some approaches
which conflate doing a query with also signaling an arrival (i.e., only query
about whether an arrival is a duplicate) we consider the general case, in which
queries and arrivals (insertions) are independent. This is useful, e.g., when
generating the stream, to test whether each of several candidates for the next
element would be a duplicate, if chosen. The properties to be respected by an
implementation of an $(w, s, p, e)$-Sliding Filter are:

\begin{description}
\item[No false negatives] If an element $x$ was added in the last $w$
    insertions, $\mathsf{query}(x)$ must return true;
  \item[Slack] If an element $x$  was inserted within $s$ insertions before
    the last $w$ insertions, $\mathsf{query}(x)$ can return indifferently
    either true or false;
  \item[Probability-weighted slack] The expected number of elements, inserted
    within $s$ insertions before the last $w$ insertions, for which the query
    returns true is less or equal to $p$;
  \item[False positives bound] Queries for elements never inserted, or
    inserted before the last $w+s$ insertions, can return true with
    probability at most $e$.
\end{description}

The definition of element slack characterizes how fast an element is forgotten and falls into the general false positive background, it allows for $s$ elements not included in the last $w$ to still be remembered in the filter. This value will depend of the garbage collection approach of each solution, and ideally should be low. 

%% file: related-work.tex
\section{Bloom Filter based approaches for duplicate detection in streams}
\label{sec:related}


Existing probabilistic data structures for duplicate detection on streams can be
broadly divided into Bloom filter based and Dictionary-based (even if the ``Bloom
filter'' term is sometimes abused, with the risk of losing its meaning).
Essentially, Bloom filter variants use $k$ hash functions to choose $k$ cells
to update, each holding some bit or counter or timestamp; cells from
insertions of different elements end up mixed (e.g., ORed or ADDed). Examples
are Counting Bloom Filters~\cite{BonomiMPSV06} or Generalized Bloom Filters~\cite{LauferVD11}.

Dictionary-based approaches use one (or a few) hash functions to choose one
cell (possibly one of a few alternatives, e.g., one of several slots in a
bucket in one of two arrays) where some content (e.g., a fingerprint and
a timestamp) will be stored. Each cell is kept separate and contents of cells
relative to different elements will not be mixed. Basically,
they are some hash-table variant based, but storing some hashes and not the
full elements themselves. Examples are Cuckoo Filters~\cite{FanAKM14} and Morton
Filters~\cite{BreslowJ18}.

Most approaches for duplicate detection in streams are BF-based but the best
ones tend to be dictionary-based, like in~\cite{NaorY15} using a Backyard
Cuckoo Hashing dictionary, or in~\cite{LiuCG13} also based on Cuckoo hashing,
or SWAMP~\cite{AssafBEF18}, using combination of dictionary mapping
fingerprints to counters plus a circular buffer of fingerprints.

Our approach improves the state of the art of BF-based approaches and contrary to
most of them is competitive space-wise with dictionary-based approaches, while
being very simple to implement. It also provides an adjacent transition zone with
exponential forgetting, outside the desired window (while garanteeing no
false-negatives inside it), a feature that may be interesting in itself, leading
to a small probability-weighted slack.

We now briefly survey BF-based approaches, to give an understanding of the
design space where our approach lies. Broadly, we can classify them as either
\emph{bit decaying based}, \emph{segmentation based}, \emph{counter based},
and \emph{timestamping based}.

\subsection{Bit decaying based approaches}

Bit decaying based approaches forget the past by resetting bits,
randomly or by further hashing elements inserted, to limit the BF fill-rate.
The big drawback is that they only \emph{tend} to forget the distant past, but
there are no guarantees that a recent element is never affected, leading to
false negatives. They tend to cause a large variance in the age when an
inserted element stops being reported, and therefore are not well suited to
the problem, in either the sliding or jumping window models.

One example is the \emph{Scope Decay Bloom Filter (SDBF)}~\cite{LiWX06}, which resets
random bits, either with an exponential decay model by resetting each bit with
a given probability (inpractical), or by a linear decay model, resetting a few
random bits each time. Another is the \emph{Generalized Bloom Filter
(GBF)}~\cite{LauferVD11}, which at each insertion uses also another set of
$k_0$ hash functions to reset $k_0$ bits. \emph{Reservoir Sampling based Bloom
Filter (RSBF)}~\cite{DuttaBN12}, a partitioned BF scheme which insert missing elements only
with some probability and if inserting also resets one random bit from each partition.
In~\cite{BeraDNB12} some \emph{Biased Sampling based Bloom Filter (BSBF)} variants
are described, such as a variant of RSFB which always inserts, a variant which
only deletes one bit from one part, and another which stores the fill-rate of
each part and uses it for probability of bit resetting from the respective part.

\subsection{Segmentation based approaches}

Segmentation based approaches use several disjoint segments which can
be individually added and retired. The most naïf and several times mentioned
approach is a sequence of plain BFs, one per generation, adding a new one and
retiring the oldest when the one in use gets full. This is a perfect fit for
the jumping window model with one BF per sub-window. One special case of this
scheme with two BFs is the \emph{Cell Bloom filter (CEBF)}~\cite{WangZDWZ14}.
Unfortunately, to have smooth jumps or to have little slack in the sliding
window model, a sequence of more than two BFs is needed and it will become
slow and not memory efficient, as a query will need testing each BF, each one
needing a tighter false-positive rate. Most approaches try some
more sophisticated segmentation. To avoid rough jumps and keep the more recent
elements alive, current approaches write in several segments when inserting, leading
to a waste of space.

The Double Buffering concept was introduced in~\cite{ChangLF04}, using a pair
of \emph{active} and \emph{warm-up} BFs, essentially using the active for
queries and inserting in both until the warm-up is half-full, at which point
it becomes the active, the previous active is discarded and a new empty warm-up is added.
Somewhat dually, Active-Active Buffering (A2 Buffering)~\cite{Yoon10}, while
having also two BFs, named \emph{active1} and \emph{active2}, inserts only in
active1 but queries in both, with the nuance that if it is only found in
active2, the element bits are copied to active1. Compared with Double
Buffering it is more memory efficient, as both active1 and active2 can store
distict data, while in that scheme one BF is always a subset of the other.

Forgetful Bloom Filter (FBF)~\cite{SubramanyamGLW16}, uses a \emph{future}, a
\emph{present} and $N \geq 1$ \emph{past} BFs. Inserts in the future and
present components and queries (essentially) by testing the presence in two
consecutive BFs from future to oldest past. When full, the oldest past is
discarded, other segments are shifted and a new future is added. The waste of
space by duplicate insertion is somewhat compensated by the reduced
false-positive rate through the two-consecutive filters test. However, the
correlation between consecutive segments caused by duplicate insertion leads
to the false-positive rate reduction being modest and not space efficient.

The more sophisticated Segmented Aging BF (SA-BF)~\cite{KaoLCW18} combines the
active/warm-up approach with a partitioned scheme, being each segment a
partitioned BF with $k$ parts. Insertions go to both active and warm-up but
only the active is queried. At forget time, one part is moved from the warm-up
to the active, in a round-robin scheme over the $k$ parts. Regardless of the
sophistication, the duplicated insertion causes some space inefficiency.

\subsection{Counter based approaches}

Many approaches are based on counting BFs~\cite{BonomiMPSV06}, using the same
representation (a vector of counters), not merely to allow element deletion in
sets, but for other purposes such as representing multi-sets.
One example is the above mentioned~\cite{MetwallyAA05}, which for the sliding
window model uses a counting BF to store a multi-set, counting the number of
occurrences in the window. While counting BFs are relatively efficient for
their original purpose (allowing deletion in a set), as it is enough to have 4
bit counters, if used to store multi-sets the size of each counter will render
them very inefficient when compared with dictionary-based approaches.

Also, several approaches, as the one just mentioned, require knowing the elements
themselves to expire them when they leave the window, using a window sized
circular queue of elements, aggravating the space consumption problem. This is
the case when using \emph{Spectral Bloom Filter}~\cite{CohenM03}, or Floating
Counter Bloom Filter (FCBF)~\cite{WangS10}, this last using even more space by
having floating point numbers, aimed at reporting \emph{existential
probabilities}.

Most counter based approaches avoid storing elements themselves by storing a fixed
value in each of the $k$ cells when inserting, and periodically decrement
counters, considering an element absent if one or more counters has reached
zero. A basic approach would be using the window size $N$ as starting value
and decrementing all counters in the filter per insertion, as in
\emph{Decaying Bloom Filters (DBF)}~\cite{ShenZ08} is unacceptably
inefficient both time and space wise; it is somewhat more acceptable if time-based
windows are used~\cite{KoloniariNPS11}, if many events fit in a time unit.
Space consumption in DBFs is addressed in the same paper by grouping elements in
generations, obtaining \emph{block Decaying Bloom Filter (b\_DBF)}; time
complexity is improved at the cost of using oversized counters, to allow less
periodical subtractions, and computing bursts avoided by de-amortizing such
subtractions over time. Similar approaches are used in the \emph{Temporal
Counting Bloom Filter (TCBF)}~\cite{ZhaoW14}, 

Unlike most counter based approaches, which guarantee no false-negatives,
\emph{Stable Bloom Filters}~\cite{DengR06} use counters but are more related to
bit decaying approaches. They decrement some random counters (if non-zero) and
set $k$ counters to some fixed value at each insertion. They allow
false-negatives and do not provide control over the expiration age of inserted
elements.

\subsection{Timestamping based approaches}

A slight variation on using counters is using integers that will remain
immutable until expiration, representing the insertion timestamp. These aim to
avoid the periodic decrementing over time of counter based approaches.

\emph{Timing Bloom Filters}~\cite{ZhangG08} stores in each of the $k$ cells the insertion
timestamp, and increases a current time variable; query is by comparing the
minimum timestamp in the $k$ cells to the current time. To have relatively
small integers, time and timestamps are stored modulo some number greater than
the window size. Using, e.g., twice the window size allows just a few elements
to be scanned for expiration at each insert.
A similar approach is used in \emph{Time Bloom Filters}~\cite{LeeC11}, which
also introduces \emph{Time Interval Bloom Filters}, which improve the
false positive rate by storing start-end timestamp intervals.

Detached Counting Bloom filter Array (DCBA)~\cite{WeiJZFW11}, addresses the
(precise) sliding window model while having a segmented architecture with a
component per sub-window, using a mix of bits and timestamps. Uses a number of
sub-windows on the order of the word size (e.g., 32, or 64), and to each sub-window 
devotes a filter with precise timestamps ranging over the window, saving in
bits per timestamp. For the inner sub-windows (that are not suffering
insertion or expiration) it groups the bits for all sub-windows for a given
hash position in the same word, to allow efficient queries in $k$ iterations
(a similar scheme, \emph{Group Bloom Filters}, was also proposed in~\cite{ZhangG08}).

Finally, an inferential version of Timing Bloom Filter~\cite{DautrichR19},
allows more sophisticated queries such as inferring the most likely insertion age of
a given element (and not merely if it is a duplicate).

\subsection{Discussion}

Considering the the Sliding Filter problem, the bit decaying approaches are
clearly inappropriate, both due to the false negatives and the little control
over expiration. Both counter based and timestamping based approaches are not
efficient in terms of memory, as they take more space than a classic counting
Bloom filter; they are not able to compete with dictionary-based approaches.

Current segmentation based approaches address slack by updating
several segments when inserting, to cause some overlap between them.
But they perform duplicate insertion, using the same hashes, which causes
memory inneficiency. Our approach, presented below, is the first segmented
based approach which does not perform duplicate writing, but uses different
hash functions to write different patterns in different segments, which will
be expired at different times.